\documentclass[aps,prb,onecolumn,11pt,a4paper,citeautoscript,floatfix]{revtex4-1}

\usepackage{amsmath,amssymb}
\usepackage{bm}
\usepackage{graphicx}
\usepackage[usenames,dvipsnames,svgnames,table]{xcolor}

\newcommand{\figbf}[1]{{\footnotesize\sffamily\bfseries #1}}
\newcommand{\La}{LaAlO$_3$}
\newcommand{\Sr}{SrTiO$_3$}

\begin{document}

\title{Long-range electronic reconstruction to a $d_{xz,yz}$-dominated Fermi surface\\below the LaAlO$_3$/SrTiO$_3$ interface}

\author{A.P. Petrovi\'c$^1$$^,$$^\dagger$, A. Par\'e$^1$, T.R. Paudel$^2$, K. Lee$^1$$^,$$^3$, S. Holmes$^4$, C.H.W. Barnes$^3$, A. David$^1$, T. Wu$^1$, E.Y. Tsymbal$^2$ and C. Panagopoulos$^1$$^,$$^3$$^,$$^\dagger$} 

\affiliation{{$^{1}$\mbox{School of Physical and Mathematical Sciences, Division of Physics and Applied Physics,} Nanyang Technological University, 637371 Singapore}\\
{$^{2}$Department of Physics and Astronomy, Nebraska Center for Materials and Nanoscience, University of Nebraska Lincoln, Nebraska 68588-0299, USA}\\
{$^{3}$Cavendish Laboratory, University of Cambridge, Cambridge CB3 0HE, United Kingdom}\\
{$^{4}$Toshiba Research Europe Ltd., Cambridge Research Laboratory, 208 Cambridge Science Park, Milton Road, Cambridge CB4 0GZ, United Kingdom}
%{$^\dagger$Corresponding Author}
}
\date{\today}

\begin{abstract}
Low dimensionality, broken symmetry and easily-modulated carrier concentrations provoke novel electronic phase emergence at oxide interfaces.  However, the spatial extent of such reconstructions - i.e. the interfacial ``depth'' - remains unclear.  Examining {\La}/{\Sr} heterostructures at previously unexplored carrier densities $n_{2D}\geq6.9\times10^{14}$~cm$^{-2}$, we observe a Shubnikov-de Haas effect for small in-plane fields, characteristic of an anisotropic 3D Fermi surface with preferential $d_{xz,yz}$ orbital occupancy extending over at least 100~nm perpendicular to the interface.  Quantum oscillations from the 3D Fermi surface of bulk doped {\Sr} emerge simultaneously at higher $n_{2D}$.  We distinguish three areas in doped perovskite heterostructures: narrow (\,$<20$~nm) 2D interfaces housing superconductivity and/or other emergent phases, electronically isotropic regions far (\,$>120$~nm) from the interface and new intermediate zones where interfacial proximity renormalises the electronic structure relative to the bulk.  

\end{abstract}

\maketitle

\let\thefootnote\relax\footnotetext{\scriptsize{{$^\dagger$}Correspondence and requests for materials should be addressed to A.P.P. (email: appetrovic@ntu.edu.sg) or C.P. (email: christos@ntu.edu.sg)}}

Ever since the discovery of a conducting channel in {\La}/{\Sr}~\cite{Ohtomo-2004} and the subsequent observations of magnetism~\cite{Brinkman-2007} and superconductivity~\cite{Reyren-2007}, the vast majority of oxide interface research has focussed on synthesising intrinsically-doped heterostructures featuring narrow conducting channels ($\lesssim$~20~nm) with two-dimensional carrier densities $n_{2D}$ in the 10$^{12}$-10$^{14}$~cm$^{-2}$ range~\cite{Salluzzo-2009,Caviglia-2008,Bell-2009,Caviglia-2010,Shalom-2010,Chambers-2010}.  At such interfaces, it has been shown \cite{Popovic-2008,Pentcheva-2010,Delugas-2011,Khalsa-2012,Salluzzo-2009,Santander-2011} that symmetry-lowering and quantum confinement lift the Ti $t_{\mathrm{2g}}$ degeneracy, so that the $d_{xy}$ orbital lies at lower energy than the $d_{xz,yz}$ orbitals.  X-ray absorption spectroscopy~\cite{Salluzzo-2009} reveals a band splitting of $\sim$~50~meV for $n_{2D}\sim$~10$^{13}$~cm$^{-2}$ and theoretical approaches indicate that this increases with $n_{2D}$, reaching $\sim$~0.25~eV at 3$\times$10$^{14}$~cm$^{-2}$~\cite{Delugas-2011}.  Regardless of the total $n_{2D}$, the splitting should gradually vanish below the interface, until the electronic structure resembles that of bulk {\Sr} with degenerate $d_{xy,xz,yz}$ orbitals at the centre of the Brillouin zone~\cite{Marel-2011}.  The lengthscale over which this degeneracy is regained - i.e. the total distance over which the interface induces electronic reconstruction - remains unknown, despite being a vital prerequisite for building layered 3D oxide devices.  

Probing this lengthscale requires the synthesis of {\La}/{\Sr} heterostructures with significantly more carriers (and correspondingly deeper conducting channels) than the norm.  Previously, high $n_{2D}$ heterostructures have only been grown in reducing environments~\cite{Herranz-2007,Basletic-2008}, creating bulk-like conducting layers hundreds of microns thick ($n_{2D}\geq$~10$^{16}$~cm$^{-2}$) in which the broken symmetry of the interface plays no role.  However, interfaces with 5$\times$10$^{14}~{\lesssim}~n_{2D}\lesssim$~5$\times$10$^{15}$~cm$^{-2}$ have until now remained unexplored: at these intermediate $n_{2D}$, electrons ``spill over'' from the interface and begin to occupy states lying deeper within the {\Sr}.  The principal focus of our work is therefore to track the evolution of the electronic structure and its crossover from 2D interfacial to 3D bulk-like behaviour within this range of carrier densities.  For $n_{2D}\geq$~6.9$\times$10$^{14}$~cm$^{-2}$, we report the first instance of Shubnikov-de Haas (SdH) oscillations from an ultra-high mobility electron gas ($\mu_H\sim10^4$~cm$^2$V$^{-1}$s$^{-1}$) for small magnetic fields parallel to the interface.  The absence of such oscillations from the low-field perpendicular magnetoresistance indicates that these carriers originate from an anisotropic 3D Fermi surface (FS); our first-principles calculations of the sub-interfacial electronic structure reveal dominant $d_{xz,yz}$ orbital occupancy, which is consistent with our experimental data.  Superconductivity remains confined within 20~nm of the interface, while the 3D FS characteristic of bulk doped {\Sr} gradually emerges with increasing $n_{2D}$.  Together, our results imply the existence of a region below the interface whose electronic structure differs from that of the bulk, with a minimum thickness of 100~nm imposed by the cyclotron radius.  This discovery has important implications for oxide devices seeking to functionalise interfacial electronic reconstructions.

\section*{Results}

\begin{figure}[htbp]
\includegraphics*[width=0.5\columnwidth]{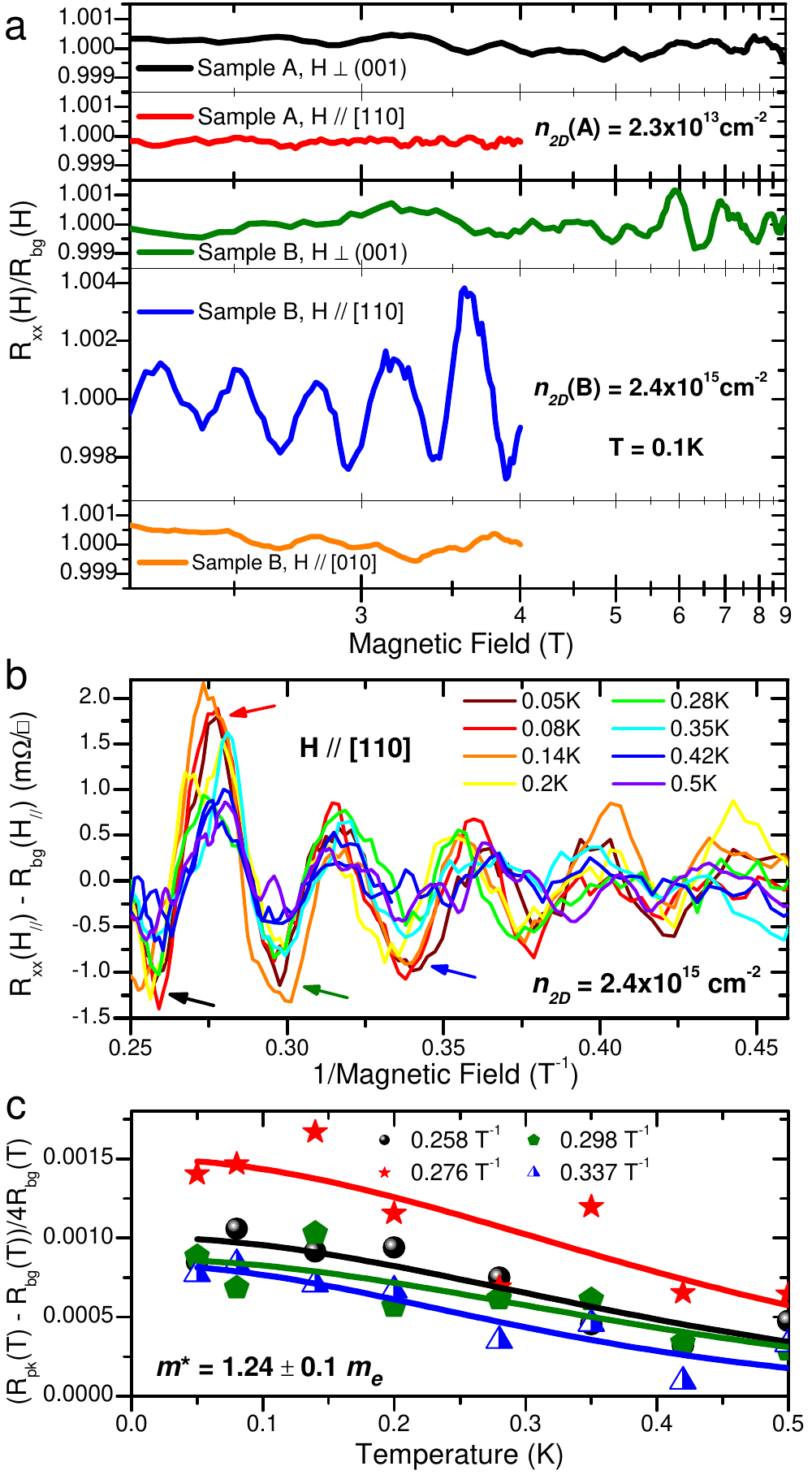}
\caption[]{\label{Fig1}
\figbf{Shubnikov-de Haas oscillations and effective mass of the high mobility electron gas below the interface.}\newline  
\figbf{a}, Oscillatory components of the MR $R_{xx}(H_{\perp,/\!/})/R_{bg}$ in samples A and B.  The background $R_{bg}$ was determined using a polynomial fit to the raw $R_{xx}(H)$ data for 1.5~T~$<H_\perp<$~9~T and 2~T~$<H_{/\!/}<$~4~T.  Data for sample B were acquired at $n_{2D}$~=~2.4$\times$10$^{15}$~cm$^{-2}$, i.e. $V_g$~=~350~V.  The vertical scale for these graphs is identical.  \figbf{b}, Temperature variation of the in-plane SdH oscillation amplitude in sample B ($V_g$~=~350~V).  \figbf{c}, Amplitude suppression with increasing temperature for four in-plane SdH MR oscillation peaks/troughs $R_{pk}(T)$  (indicated by arrows in \figbf{b}).  Symbols represent oscillation amplitude data-points, while the four colour-coordinated lines are least-squares fits using equation~\ref{Eq2} to calculate $m^*$.  Due to the low resistance of sample B ($\sim$~1~$\Omega$), the SdH effect for $H\perp(001)$ creates voltage fluctuations of $\sim$~1~nV, which is the noise threshold in our measurements.  Consequently, oscillations were only visible below $T$~=~0.1~K and $m^*$ could not be determined for these carriers.}
\end{figure}

During sample growth, three mechanisms exist for carrier-doping the {\La}/{\Sr} interface: (a) intrinsic self-doping via the polar catastrophe~\cite{Nakagawa-2006} (leading to a maximum $n_{2D}$~=~3.3$\times$10$^{14}$~cm$^{-2}$), (b) oxygen vacancy doping~\cite{Pavlenko-2012-2} (contributing 2$e^-$ per O$^{2-}$ vacancy) and (c) cation intermixing~\cite{Chambers-2010} (an unbalanced switching of La$^{3+}$ for Sr$^{2+}$ and Al$^{3+}$ for Ti$^{4+}$).  Since our principal aim is to explore the evolution of the electronic structure for $n_{2D}>$~5$\times$10$^{14}$~cm$^{-2}$ (far beyond the upper limit imposed by the polar catastrophe) and cation intermixing is difficult to control in a pulsed laser deposition (PLD) chamber, we use O$^{2-}$ vacancy doping to achieve the high $n_{2D}$ values necessary for this project.  To this end, we synthesise {\La}/{\Sr} heterostructures at an intermediate O$_2$ pressure (10$^{-3}$~mbar), without any post-annealing procedure (further growth and characterisation details may be found in the Methods and Supplementary Material).  The lack of annealing guarantees a high O$^{2-}$ vacancy concentration and hence a large $n_{2D}$, while the intermediate growth pressure ensures that these vacancies do not penetrate far into the {\Sr} substrate.  Low pressure growth (10$^{-6}$~mbar) without annealing~\cite{Herranz-2007,Basletic-2008} has previously been shown to result in macroscopic substrate conduction, with $n_{2D}\geq5\times10^{15}$~cm$^{-2}$; in contrast, our method of synthesis consistently yields heterostructures with as-grown Hall carrier densities in the 10$^{14}$-10$^{15}$~cm$^{-2}$ range, which we will refer to as ``series B''.  For comparative purposes, we have also annealed certain heterostructures (``series A'') at high $O_2$ pressures, yielding $n_{2D}\sim$~10$^{13}$~cm$^{-2}$.  A-type interfaces are comparable to the majority of those previously studied in the literature~\cite{Reyren-2007,Caviglia-2008,Bell-2009}, in which carrier injection is dominated by the polar catastrophe.  Both series exhibit coexistent superconductivity (SC) and ferromagnetism (FM), a comprehensive analysis of which may be found in ref.~\cite{Petrovic-2013}. For quantitative continuity in the present work, we focus on two specific samples A and B, with as-grown $n_{2D}$~=~2.3$\times$10$^{13}$~cm$^{-2}$, 6.9$\times$10$^{14}$~cm$^{-2}$ at $T$~=~0.1~K and SC channel thicknesses $d$~=~18$\pm$1~nm, 9$\pm$1~nm respectively.  Sample B has a back gate beneath the {\Sr}: $n_{2D}$ increases to 2.4$\times$10$^{15}$~cm$^{-2}$ ($d$~=~19$\pm$2~nm) at gate voltage $V_g$~=~350~V.  The heterostructure withstands $V_g$~=~500~V with no discernible leakage current and the substrate capacitance $\sim$~1~nF is comparable to values measured in annealed {\La}/{\Sr} heterostructures with lower $n_{2D}$~\cite{Caviglia-2008,Bell-2009} (see Supplementary section 1).  Such conditions can only be achieved if the bulk of the {\Sr} substrate is insulating: this confirms that O$^{2-}$ vacancies have not penetrated deep into the {\Sr} and are restricted to the neighbourhood of the interface.  

We probe the electronic structure and FS geometry using SdH oscillations in the magnetoresistance (MR) $R_{xx}(H)$ (Fig.~1a).  Two magnetic field orientations are principally considered: $H\perp(001)$ ($H_{\perp}$) and $H/\!/[110]$ ($H_{/\!/}$), where the $[100]$ directions correspond to the crystallographic axes of the {\Sr} substrate and $[001]$ points out-of-plane.  Sample A does not display any SdH effect for either orientation.  In contrast, sample B exhibits strong oscillations for $H_{/\!/}$ as low as 2.5~T, with faint oscillations also emerging for $H_{\perp}>$~6~T.  However, data acquired with an in-plane field $H/\!/[010]$ do not show any oscillations up to 4~T.  Symmetry dictates that the plane of a 2D FS in {\La}/{\Sr} must lie parallel to the interface; any such FS will therefore lack states with out-of-plane momenta and cannot exhibit any SdH effect for in-plane fields.  It is therefore immediately clear that the oscillations which we observe with $H/\!/[110]$ must originate from an anisotropic 3D FS.

For $H/\!/[110]$, the SdH oscillations in sample B are sufficiently pronounced for us to extract the effective mass $m^*$ and the Dingle temperature $T_D$ (a measure of the scattering) from their temperature-dependent amplitude (Fig.~1b).  The magnitude of the oscillatory resistance is given by:
\begin{multline}
\label{Eq2}
(R_{xx}(H,T)-R_{bg}(H,T))/4R_{bg}~=~\mathrm{exp}(-\frac{2\pi^2m^*k_BT_D}{e{\hbar}H})\frac{2\pi^2m^*k_BT}{e{\hbar}H}/\mathrm{sinh}(\frac{2\pi^2m^*k_BT}{e{\hbar}H})
\end{multline}
where $R_{bg}$ is the background resistance.  Fitting this equation to the oscillation amplitude (Fig.~1c) yields $m^*$~=~1.24$\pm$0.1~$m_e$ and $T_D$~=~1.4$\pm$0.4~K.  $m^*$ is similar to values previously reported for the {\La}/{\Sr} 2DEG~\cite{Caviglia-2010,Shalom-2010}, although our $T_D$ is lower which implies a higher carrier mobility in our heterostructures.  To estimate this mobility, we initially calculate the Hall mobility $\mu_H=1/n_{2D}eR_{xx}$, where $R_{xx}(V_g=0)=$~0.28~$\Omega/\square$ and we assume single-band transport.  This yields an exceptionally high Hall mobility $\mu_H=$~32000~cm$^2$V$^{-1}$s$^{-1}$, setting a new record for pure {\La}/{\Sr} and rivalling the best epitaxial {\Sr} films~\cite{Son-2010}.  

In order to justify such a high mobility, we evaluate the Drude scattering time $\tau_{dr}~{\equiv}~m^*\mu_H/e=$~23~ps, which is more than an order of magnitude greater than the Dingle scattering time $\tau_D\equiv\hbar/2{\pi}k_BT_D=$~0.87~ps.  An alternative estimate of the scattering time in sample B may be extracted from the field at which a SdH effect first appears, using the quantum oscillation emergence condition $\omega_c\tau_{SDH}\sim1$ (where $\omega_c~{\equiv}~Be/m^*$ is the cyclotron frequency and $B$ the magnetic field strength).  For $H/\!/[110]$, oscillations are visible above 2.5~T: this corresponds to $\tau_{SDH}=$~2.8~ps, which is also shorter than $\tau_{dr}$ suggested by our high $\mu_H$. It is likely that four factors contribute to this disparity: firstly, all scattering events suppress quantum oscillations and contribute to $\tau_D$, while only back-scattering influences $\tau_{dr}$ and the Drude conductivity.  Similar variance between $\tau_D$ and $\tau_{dr}$ can be seen in other {\La}/{\Sr} heterostructures~\cite{Caviglia-2010}.  Secondly, the finite thickness of the conducting channel in our heterostructures may postpone the emergence of any SdH effect, until the applied field is sufficiently large for the diameter of the cyclotron orbits to fall below this thickness.  Thirdly, superconducting fluctuations at fields below $\sim$~2.5~T effectively ``short-circuit'' our heterostructures, reducing our ability to probe transport from carriers deeper below the interface.  Finally, our single-band estimate for $\mu_H$ is an over-simplification, since multiband transport is expected for carrier densities above the Lifshitz transition in {\La}/{\Sr}~\cite{Delugas-2011,Joshua-2011}.  A three-band approximation to the field-dependent Hall coefficient (see Supplementary section 2) suggests a minority contribution from a high-mobility band with $\mu_H\approx$~8000~cm$^2$V$^{-1}$s$^{-1}$.  The total number of conduction bands in our heterostructures and their field-dependent mobilities remain unknown, so we cannot obtain a more precise value for the mobility of these quantum-oscillating carriers.  However, it is clear that our SdH effect, resistivity and Hall data all indicate the presence of a high-mobility band with an anisotropic FS and $\mu_H\sim$~10$^4$~cm$^2$V$^{-1}$s$^{-1}$.

\begin{figure}[htbp]
\includegraphics*[width=0.46\columnwidth]{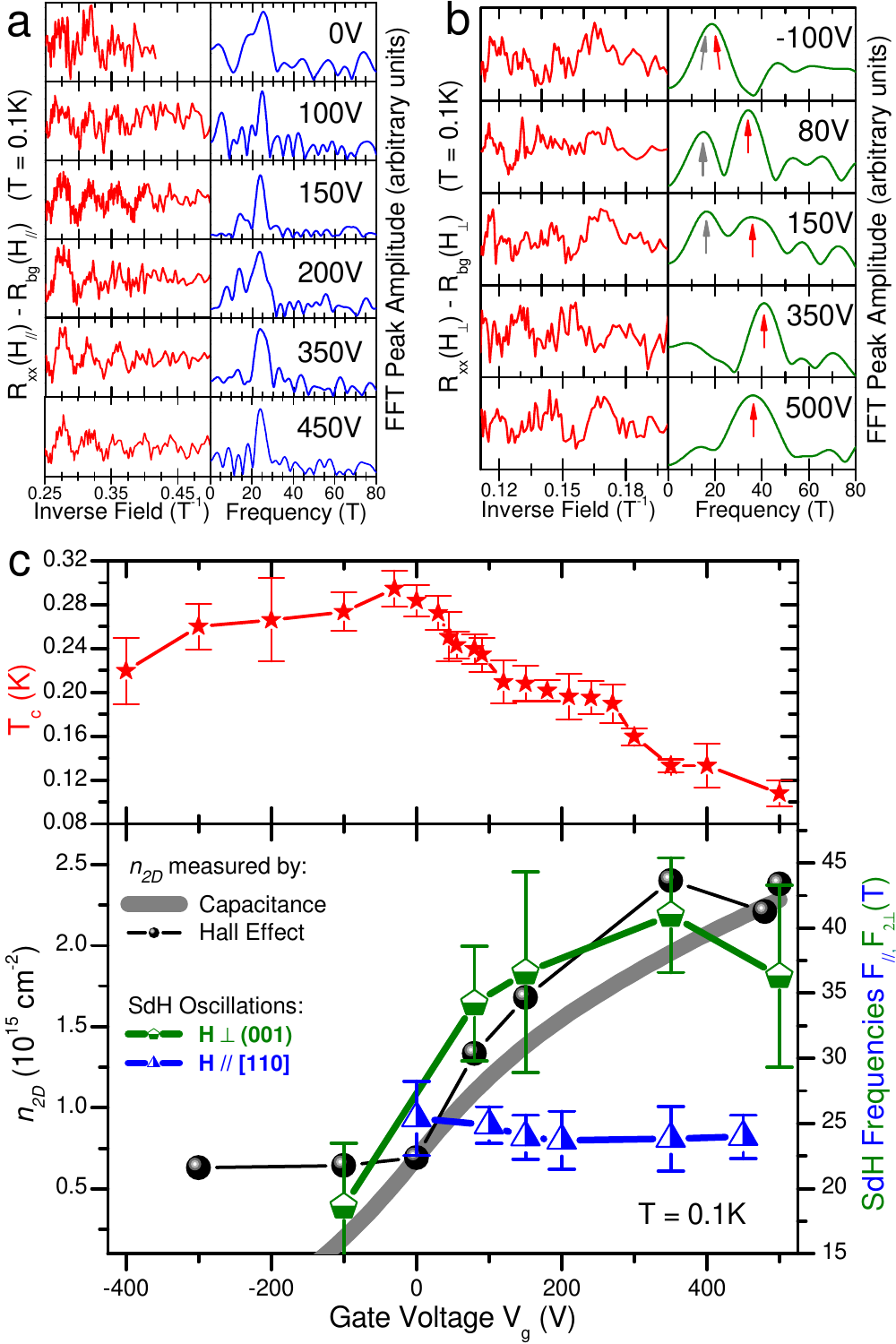}
\caption[]{\label{Fig2} 
\figbf{Evolution of the Shubnikov-de Haas oscillations and carrier density with increasing gate voltage.}\newline \figbf{a}, Variation of the oscillating component of $R_{xx}(H_{/\!/}$) for sample B (left panel) at $V_g\geq0$, with FFTs of the raw data (right panel).  For $V_g<0$ the noise level rises and it is not possible to identify oscillations: this is a well-known phenomenon and has been attributed to emergent inhomogeneity~\cite{Bell-2009}.  \figbf{b}, Oscillating components of $R_{xx}(H_\perp)$ for $V_g\geq$~-100~V in sample B (left panel) with associated FFTs (right panel).  The two peaks in the FFTs are indicated by grey and red arrows; for $V_g$~=~-100~V, the peaks merge.  \figbf{c}, $V_g$ dependence of various properties of sample B, including $T_c$ (above), SdH frequencies $F_{/\!/,2\perp}$ and $n_{2D}$ (below).  $T_c$ is measured from $R_{xx}(T)$ data (see Supplementary Fig.~S2) and the errors in $F_{/\!/,2\perp}$ correspond to the FFT peak widths at 80\% of their maximum height (from \figbf{a},\figbf{b}).  $n_{2D}(V_g)$ obtained from the Hall coefficient~\cite{Petrovic-2013} follows the values expected from the sample capacitance $C(V_g)$ (see Supplementary Fig.~S1b for raw capacitance data).  We attribute the fall in $n_{2D}$ above $V_g=350$~V to charge-trapping deep within the {\Sr}.  $T_c(V_g)$ forms a dome: since $d$~=~19~nm at $V_g$~=~350~V and {\Sr} is SC for 5.5$\times$10$^{17}$~cm$^{-3}{\leq}n_{3D}{\leq}5\times$10$^{20}$~cm$^{-3}$~\cite{Koonce-1967,Lin-2013}, we estimate a maximum conducting channel thickness $W\sim$~20~$\mu$m due to the combination of carrier injection and electron gas decompression~\cite{Bell-2009}.  In practice, we anticipate $W\lesssim$~1~$\mu$m due to the extremely high $n_{2D}$ at the interface which will locally suppress SC: even at $V_g=350$~V, every carrier in sample B could be accommodated in merely 3~nm of {\Sr} doped at 0.5$e^-$/unit cell. 
}
\end{figure}

The fact that our measured $T_D$ is lower than than those reported for the {\La}/{\Sr} 2DEG~\cite{Caviglia-2010,Shalom-2010} also suggests that the band whose FS generates the in-plane oscillations lies within an extremely clean region of our heterostructures, far from the cation defects and magnetic scattering expected at oxygen-deficient PLD-grown {\La}/{\Sr} interfaces.  To determine the location of these high-mobility carriers more precisely, we examine the evolution of the SdH oscillation frequencies with field-effect doping, obtained from the peaks in fast Fourier transforms (FFTs) of $R_{xx}(H_{/\!/,\perp})$ (Fig.~2a,b).  The Onsager relation links the peak frequency $F$ with the extremal area $S$ of the FS normal to the applied field via $F=\frac{S{\hbar}}{2{\pi}e}$: since the size of the FS should be proportional to the carrier density, it is useful to compare $F(V_g)$ with our experimentally-determined total $n_{2D}$ as well as the superconducting critical temperature $T_c$ (which varies strongly with the local three-dimensional carrier density $n_{3D}$~\cite{Koonce-1967,Lin-2013}).  Once the interfacial carrier density exceeds $n_{3D}\sim$~10$^{20}$~cm$^{-3}$, we expect a gradual suppression of SC leading to a dome in $T_c(V_g)$~\cite{Caviglia-2008,Bell-2009}; this is indeed observed (Fig.~2c).  However, the in-plane oscillation frequency $F_{/\!/}$ is independent of $V_g$, implying that the FS area $S\!\perp\!(110)$ responsible for these oscillations remains roughly constant upon field-effect doping.  Furthermore, $F_{/\!/}(V_g)$ displays no correlation with $T_c(V_g)$ or $n_{2D}(V_g)$: the FS (and hence the density of states) of the SC band(s) is being influenced by field-effect doping, but the FS of the high-mobility band is not.  Field-effect doping should have a similar effect on all occupied bands within the same spatial region.  Therefore, the only possible explanation for this decoupling between $T_c(V_g)$ and $F_{/\!/}(V_g)$ is that the SdH-oscillating electron gas must be spatially separated from superconductivity, i.e. the high-mobility carriers lie below the SC channel.

The gate evolution of $R_{xx}(H_\perp)$ is very different from $R_{xx}(H_{/\!/})$, with two $V_g$-dependent peaks appearing in the FFTs (Fig.~2b).  One of these ($F_{1\perp}$, grey arrows) lies below 20~T and is suppressed for large $V_g$: although this frequency seems too low to originate from the $d_{xy}$ interfacial bands (which form a larger FS at much lower $n_{2D}$~\cite{Caviglia-2010,Shalom-2010}), spin-orbit splitting may create a series of small FS for high $n_{2D}$ at the interface.  The other peak ($F_{2\perp}$, red arrows) mirrors $n_{2D}(V_g)$ as $V_g$ increases, saturating and broadening at $\sim$~40~T for large $V_g$.  This implies that $F_{2\perp}$ also cannot arise from a $d_{xy}$ 2DEG at the interface, since for back-gate doping at $n_{2D}\gtrsim5\times$10$^{14}$~cm$^{-2}$ the interfacial $d_{xy}$ occupancy should not change significantly: instead, carriers move deeper into the {\Sr}.  It is therefore tempting to link this peak with the 40~T mode from de Haas-van Alphen experiments~\cite{Gregory-1979} on $\delta$-doped bulk {\Sr}; however the light 3D band whose spherical FS was shown to be responsible for the 40~T oscillation~\cite{Marel-2011} is only occupied for $n_{3D}>$~6.7$\times$10$^{17}$~cm$^{-3}$, by when {\Sr} already shows SC~\cite{Lin-2013}.  Since $d\leq$~20~nm for sample B~\cite{Petrovic-2013}, $n_{3D}\leq$~5.5$\times$10$^{17}$~cm$^{-3}$ below the SC channel, ruling out any occupancy of this light band.  We therefore attribute $F_{2\perp}$ to the gradual population of the 3D FS from the first occupied band in bulk doped {\Sr}, which is formed by degenerate Ti 3$d_{xy,xz,yz}$ orbitals and remains approximately isotropic for such low $n_{3D}\lesssim10^{17}$~cm$^{-3}$.  

It is clear that the in-plane SdH effect $F_{/\!/}$ in our data is unrelated not only to $F_{2\perp}$, but also to any previously reported 2D~\cite{Caviglia-2010,Shalom-2010} or 3D~\cite{Herranz-2007} quantum oscillations in {\La}/{\Sr}.  Instead, our oscillations originate from a highly anisotropic FS (since there is no SdH effect for $H_{\perp}<$~6~T), which occupies a clean intermediate region between the interface and the bulk.  We estimate the minimum thickness of this region using the cyclotron radius $r_g=\frac{{\hbar}k_F}{eB}$: since a depth of at least 2$r_g$ is necessary to establish SdH oscillations for $H/\!/(001)$, we use $k_F$~=~$\sqrt{2{\pi}F_{/\!/}/\Phi_0}$ (where $\Phi_0$ is the magnetic flux quantum and we assume a spherical FS for simplicity), obtaining 2$r_g\sim$~140~nm at 2.5~T.  

\begin{figure}[htbp]
\includegraphics*[trim=1cm 10cm 0cm 6cm, clip=true, width=0.95\columnwidth]{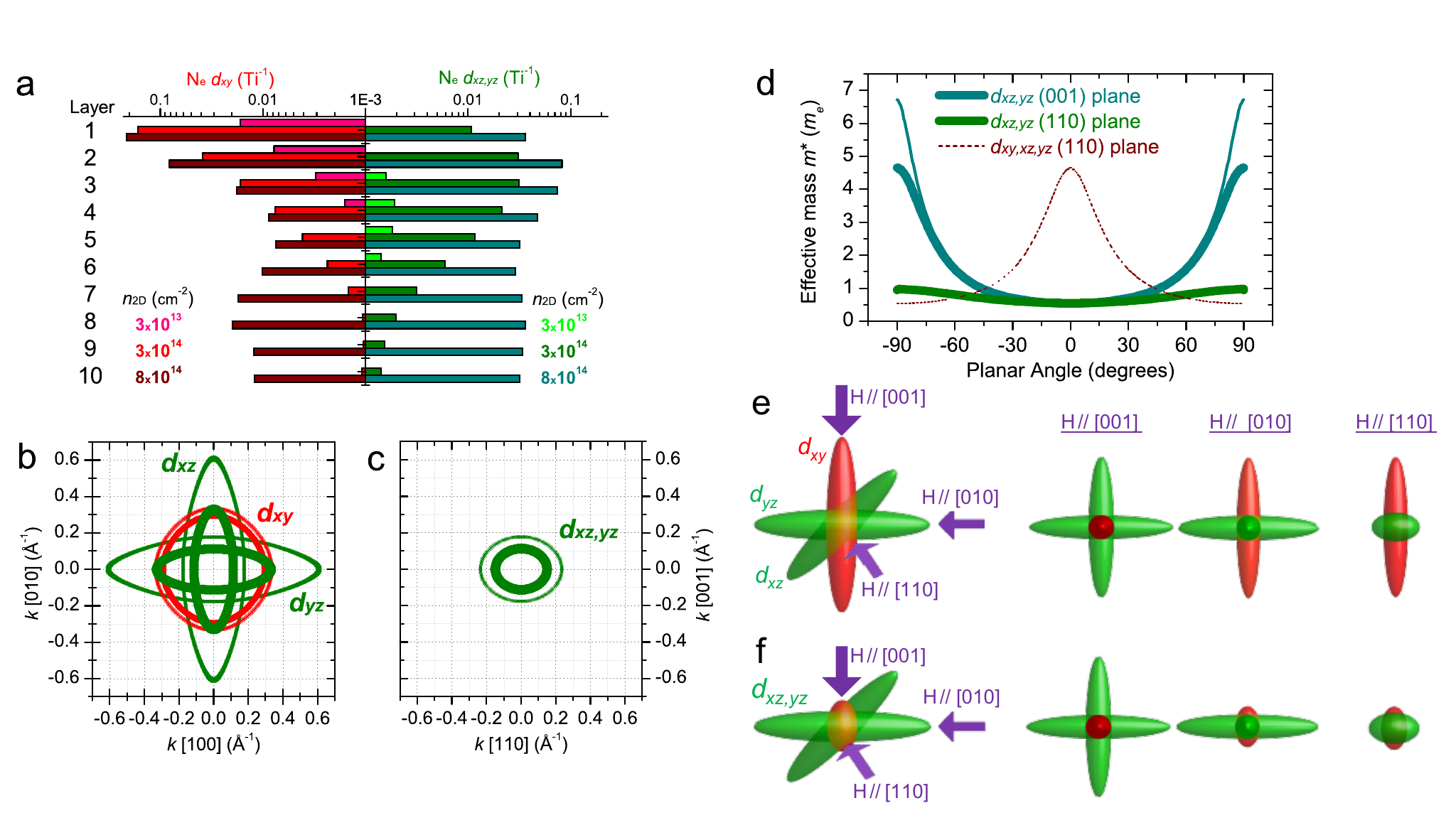}
\caption[]{\label{Fig3}
\figbf{Orbital occupancy and Fermiology calculations at high carrier densities in {\La}/{\Sr}.}\newline
\figbf{a}, Layer-dependent orbital populations for $n_{2D}$~=~3$\times$10$^{13}$, 3$\times$10$^{14}$ and 8$\times$10$^{14}$~cm$^{-2}$.  Data are plotted on a logarithmic scale with a lower cut-off of 10$^{-3}$~electrons per Ti atom.  \figbf{b},\figbf{c}, Fermi surfaces of the interfacial $d_{xy}$ (red) and $d_{xz,yz}$ bands (green) projected onto the (001) and (110) planes for $n_{2D}$~=~8$\times$10$^{14}$~cm$^{-2}$ (thin lines) and 3$\times$10$^{14}$~cm$^{-2}$ (thick lines).  The FS are calculated using a tight-binding model in which the parameters are fitted to bands calculated from first principles (see Supplementary Material for the band structure).  For simplicity, our diagram ignores the hybridization which lifts the degeneracy at the band crossing points; taking this into account would split the doubly-degenerate FS into two.  \figbf{d}, Effective mass variation $m^*_b(k_F)$ for $d_{xz}$ electrons within the (001) and (110) planes: thin and thick lines denote results at $n_{2D}$~=~8$\times$10$^{14}$ and 3$\times$10$^{14}$~cm$^{-2}$.  The planar angles are measured from the [100] and [001] axes respectively.  $m^*_b(k_F)$ for $d_{yz}$ electrons in the (001) plane is equivalent to $m^*_b(k_F)$ for the $d_{xz}$ FS rotated by 90$^\circ$ around [001].  A sketch of the mass variation expected in the (110) plane for a bulk degenerate $d_{xy,xz,yz}$ FS is also shown (dashed brown line).  \figbf{e,f}, Schematics illustrating the extremal FS orbits normal to magnetic fields along [001], [010] and [110] for a degenerate $d_{xy,xz,yz}$ FS (\figbf{e}) and a $d_{xz,yz}$-dominated FS (\figbf{f}).  Only $H/\!/[110]$ in the $d_{xz,yz}$-dominated case probes the small, light FS cross-section whose presence we infer from our in-plane SdH oscillations.
}
\end{figure}
 
To understand the origin of these in-plane SdH oscillations, we calculate the evolution of the sub-interfacial orbital occupancy (which determines the FS symmetry) with increasing $n_{2D}$.  The majority of electronic structure calculations for {\La}/{\Sr} to date have only considered the first few layers below the interface for $n_{2D}\leq$~10$^{14}$~cm$^{-2}$ and are of limited use in our heterostructures.  We have therefore performed first-principles calculations of the depth-dependent band structure in {\La}/{\Sr} for $n_{2D}$~=~3$\times$10$^{13}$, 3$\times$10$^{14}$ and 8$\times$10$^{14}$~cm$^{-2}$, specifically chosen to approach our experimental $n_{2D}$ in samples A, B ($V_g\sim$~0) and B ($V_g>$~0) respectively.  Our calculated orbital occupancies are plotted in Fig.~3a and can also be seen in Fig.~4a-c: although computational power limits us to considering the first 10 unit cells below the interface, this is already sufficient to reveal the FS anisotropy responsible for our in-plane SdH effect.  

The central result from these calculations is a crossover from $d_{xy}$ to $d_{xz,yz}$ occupancy as we move away from the interface.  Close to the interface and for small $n_{2D}$, $d_{xy}$ states dominate due to quantum confinement, as expected~\cite{Popovic-2008,Delugas-2011}.  The absence of a clear SdH signal from the 2D $d_{xy}$ interfacial FS in sample A is due to scattering from local moments~\cite{Popovic-2008} and the large Rashba spin-orbit coupling; we note that there are no reports of a 2D SdH effect in FM {\La}/{\Sr} in the literature.  The important new result from our calculations is the creation of a conducting ``tail'' deeper below the interface for large $n_{2D}$, with a disproportionate occupation of $d_{xz,yz}$ orbitals.  For example, the $d_{xz,yz}$:$d_{xy}$ ratio in layer 9 for $n_{2D}$~=~8$\times$10$^{14}$~cm$^{-2}$ is 2.8:1, significantly greater than the 2:1 expected in bulk {\Sr}.  A recent study of top-gated {\Sr} also hints at a low density ``tail'' of carriers persisting over at least 50 TiO$_2$ layers, independently of the total $n_{2D}$~\cite{Khalsa-2012}.  While the majority of carriers occupy tightly-bound bands close to the interface, the back-gate field in our sample B should reduce the quantum confinement and expand the ``tail'' still further into the {\Sr}: this competition between confinement and decompression is responsible for the weak variation of $F_{/\!/}(V_g)$ (Fig.~2c).  We therefore identify a $d_{xz,yz}$-dominated FS as the source of our in-plane SdH effect.  

The strong asymmetry in our observed SdH effect (i.e. the absence of oscillations for small $H_\perp$) may be explained by considering the FS geometry.  In Fig.~3b,c, we plot the calculated (001) and (110) extremal cross-sections of the interfacial FS at $n_{2D}$~=~3$\times$10$^{14}$ and 8$\times$10$^{14}$~cm$^{-2}$.  The elliptical cross-section of the $d_{xz,yz}$ FS implies that our previously-calculated $r_g$ will be scaled by $k_{F[001]}/k_{F[110]}=$~0.73, reducing the minimum thickness over which the electronic structure deviates from that of bulk {\Sr} to $\sim$~100~nm.  Furthermore, the variation in $\left|k_F\right|$ across the FS drives a corresponding modulation in the effective band mass $m^*_b$, shown for the (001) and (110) planes in Fig.~3d.  Electrons in the (001) plane are significantly heavier and hence more easily scattered: therefore, SdH oscillations will only emerge for $H_{\perp}{\gg}H_{/\!/}$.  Our measured $F_{/\!/}\sim$~25~T is clearly too small to originate from the large interfacial FS projections in Fig.~3c: instead, our in-plane oscillations are generated by a similarly-shaped smaller FS deeper below the interface, where $n_{3D}$ is lower.  The overall symmetry of the $d_{xz,yz}$ FS does not vary significantly with depth and hence our effective mass argument justifying the suppression of oscillations for $H_\perp$ remains valid.  In the (110) plane, the average band mass of the carriers is $m^*_b~=~2(m^*_{b[110]}m^*_{b[001]})/(m^*_{b[110]}+m^*_{b[001]})$~=~0.7~$m_e$, which only allows for a small electron-phonon coupling $\lambda~\sim$~0.8 when compared with our measured $m^*$~=~1.24~$m_e$ (since $m^*_b~=~(1+\lambda)m^*$).  However, we note that SdH experiments on both {\La}/{\Sr} and $n$-type {\Sr} heterostructures have persistently yielded small effective masses~\cite{Caviglia-2010,Shalom-2010,Kim-2011}. 

Identifying the role of $m^*_b(k_F)$ in determining the emergence of SdH oscillations allows us to make a profound statement regarding the shape of the in-plane oscillating FS.  In Fig.~3d, we sketch the approximate $m^*_b$ dependence in the (110) plane expected for a degenerate (bulk-like) $d_{xy,xz,yz}$ FS.  Here, the $m^*_b$ variation is similar to that in the (001) plane, though with a 180$^\circ$ rather than 90$^\circ$ period.  We attribute the absence of oscillations for small $H_\perp$ to the presence of heavy carriers in the (001) plane: therefore, the emergence of oscillations at small $H_{/\!/}$ implies that $m^*_b$ cannot rise significantly at 0$^\circ$.  Consequentially, the FS within this $\gtrsim$~100~nm sub-interfacial region must be flattened along the $[001]$ direction in comparison with the bulk, i.e. the $d_{xy,xz,yz}$ degeneracy is lifted and the $d_{xz,yz}$ orbitals are shifted to lower energy.  To illustrate this point further, in Fig.~\ref{Fig3}e,f we sketch $d_{xy,xz,yz}$ and $d_{xz,yz}$-dominated Fermi surfaces, comparing the shapes of their extremal orbits perpendicular to [001], [010] and [110].  The low-frequency SdH oscillations which we observe with $H/\!/[110]$ must originate from a FS whose extremal orbits are composed exclusively of light carriers (i.e. the FS cross-sectional area must be small): it is clear that this condition is only satisfied for the $d_{xz,yz}$-dominated FS.

\section*{Discussion}

\begin{figure}[htbp]
\includegraphics*[trim=0cm 3.6cm 0cm 0cm, clip=true, width=0.95\columnwidth]{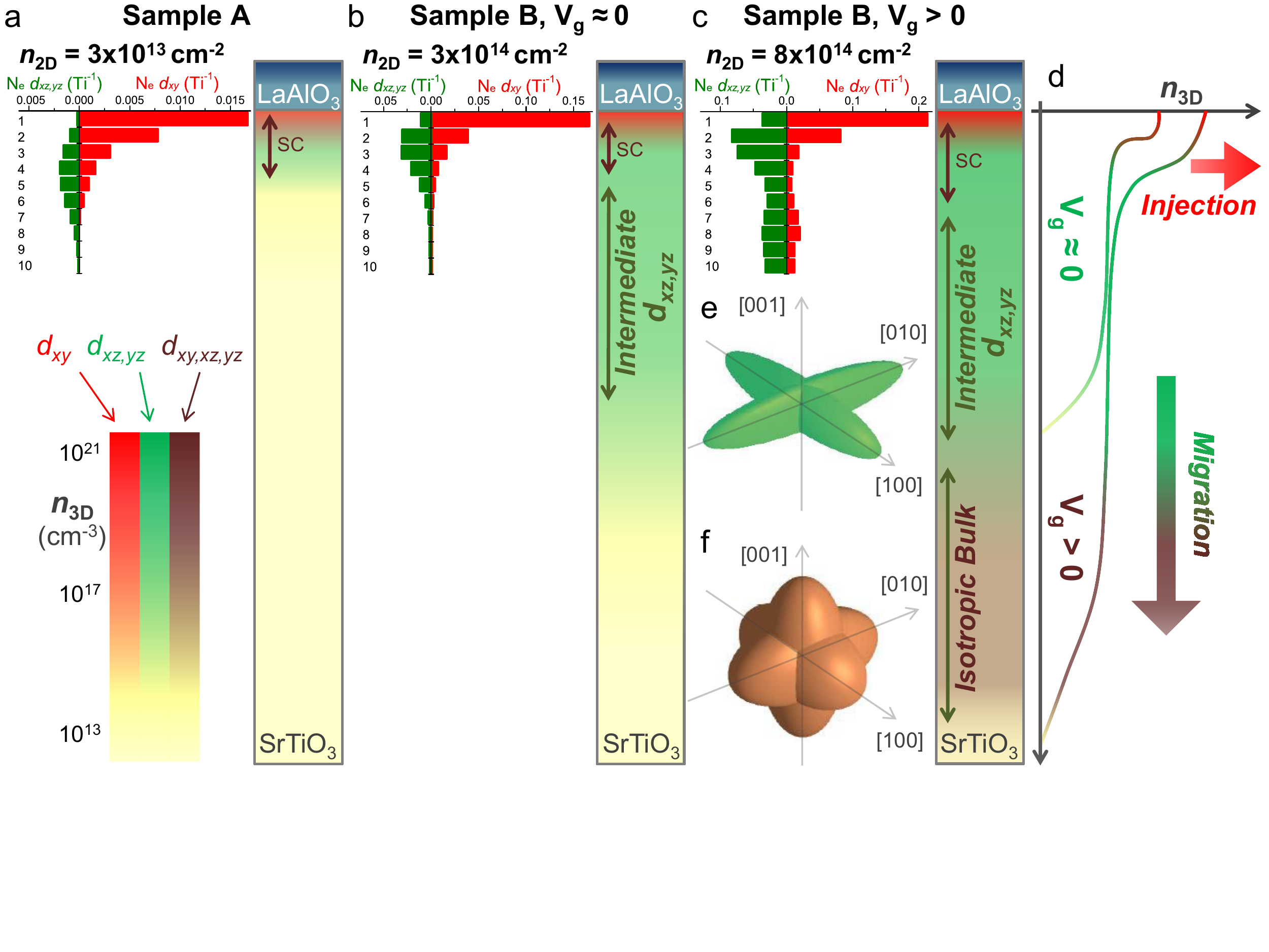}
\caption[]{\label{Fig4}
\figbf{Electronic structure, orbital character and depth-dependent phase emergence at the {\La}/{\Sr} interface.}\newline  \figbf{a-c}, Schematics illustrating the phase and carrier distributions for sample A (\figbf{a}), B at low $V_g$ (\figbf{b}) and B at large $V_g$ (\figbf{c}).  The approximate local carrier density within the {\Sr} is indicated by the colour shading: pale yellow denotes undoped bulk {\Sr}, while higher carrier density regions are either red ($d_{xy}$), green ($d_{xz,yz}$) or brown (degenerate $d_{xy,xz,yz}$) depending on the dominant orbital character.  $d_{xy}$ ferromagnetism is present regardless of the carrier density, but remains tightly confined to the interface~\cite{Petrovic-2013}, i.e. at the top of the red zone.  For comparison, we also plot the calculated orbital occupancies for the first ten TiO$_2$ layers below the interface at $n_{2D}$~=~3$\times$10$^{13}$ (\figbf{a}), 3$\times$10$^{14}$ (\figbf{b}) and 8$\times$10$^{14}$~cm$^{-2}$ (\figbf{c}).  The index ``1'' refers to the TiO$_2$ layer closest to the interface.  Red and green bars correspond to $d_{xy}$ and $d_{xz,yz}$ orbitals respectively.  \figbf{d}, Qualitative illustration of the depth-dependent influence of field-effect doping on the local carrier density, which we deduce from our transport data.  The line colour indicates the variation in the dominant orbital character with depth.  \figbf{e},\figbf{f}, Cartoon Fermi surfaces of the sub-interfacial high-mobility $d_{xz,yz}$ electron gas (\figbf{e}) and the bulk doped {\Sr} (\figbf{f}) which develops progressively for large $n_{2D}$. 
}
\end{figure} 

What is the physical origin of this change in the FS?  We note that the shape of our $d_{xz,yz}$ FS is similar to that calculated by Mattheiss~\cite{Mattheiss-1972} using a crystal-field parameter $D$ which was subsequently shown to be too large~\cite{Marel-2011}.  Since $D$ is related to the tetragonal structure of {\Sr}, our renormalised electronic structure may result from strain effects at the interface - such as the compression from the {\La} layer - which are known to influence the 2DEG~\cite{Bark-2011}.  Studies of the 2D-3D crossover in $\delta$-doped {\Sr} films~\cite{Kim-2011} (in which strain should be absent) have not revealed the $d_{xz,yz}$-dominated intermediate FS which we observe; nevertheless it remains unclear whether a long-range interface-induced change in $D$ or the spin-orbit coupling is responsible for our results.  Finally, our determination of the FS orbital character assumes the {\Sr} tetragonal $c$-axis lies parallel to $[001]$: since orthogonal tetragonal domains are expected for $T<$~105~K, this may not initially seem plausible.  However, an offset surface potential exists between domains with $c/\!/[001]$ and $c/\!/[100]$ in {\La}/{\Sr}, requiring substantial charge transfer to equalise the chemical potential~\cite{Honig-2013}.  This increases the carrier density in domains with $c/\!/[001]$, so transport predominantly occurs within these regions.  Previous transport studies of La-doped {\Sr} have also indicated a prevalence of $[001]$-oriented domains~\cite{Allen-2013}.  

We summarise the evolution of the {\La}/{\Sr} interface with $n_{2D}$ in Fig.~4, where we schematically represent the spatial distribution of SC together with the approximate $n_{3D}$ variation and our calculated depth-dependent $d_{xy}$ and $d_{xz,yz}$ orbital occupancies (Fig.~4a-c).  At low carrier densities (Fig.~4a), $d_{xy}$ orbitals dominate and the charge is concentrated within a few unit cells of the interface.  Electrons in the top TiO$_2$ layer tend to localise~\cite{Pentcheva-2006,Popovic-2008}, creating an inhomogeneous patchwork of FM zones above a narrow ($\leq$~20~nm) SC channel~\cite{Petrovic-2013}.  

As $n_{2D}$ increases (Fig.~4b), FM and SC both remain present at the interface.  However, a high-mobility $d_{xz,yz}$ ``tail'' of minimum thickness 100~nm develops below the interface, generating an anisotropic 3D FS which exhibits SdH oscillations for small in-plane fields.  Together, the appearance of this SdH effect, its independence from $n_{2D}(V_g)$ and $T_c(V_g)$, and its absence in small perpendicular fields indicate that $d_{xz,yz}$ orbital occupancy is favoured over $d_{xy}$ to a depth of at least 120~nm below the interface.  Unfortunately, it is not possible to accurately determine the maximum depth reached by this ``tail'', since the carrier density very close to the interface (where we expect the majority of the carriers to reside) is unknown.  However, our data do enable us to comment on the O$^{2-}$ vacancy penetration depth, which we already believe to be small since the capacitance of our B-type samples is comparable to values seen in annealed heterostructures.  The high electron mobility within the ``tail'' region is primarily a consequence of the low carrier density (which leads to a small FS and low effective mass), but a lack of crystal defects (e.g. O$^{2-}$ vacancies) below the interface may also play an important role.  Recently, ultra-high mobility carriers ($\mu_H\sim$~50,000~cm$^2$V$^{-1}$s$^{-1}$) have been observed in SrCuO$_2$-capped {\La}/{\Sr} heterostructures, in which O$^{2-}$ vacancy formation is suppressed~\cite{Huijben-2013}.  This suggests that although the carriers in our B-type heterostructures originate from O$^{2-}$ vacancies, these vacancies may be confined close to the interface (or in the {\La} layer) while the electrons which they donate are redistributed deeper within the {\Sr}.  This concept is supported by the absence of any parasitic {\Sr} surface conduction in our heterostructures (whose presence would be expected in the case of deep O$^{2-}$ vacancy penetration), as well as theoretical work which indicates that O$^{2-}$ vacancies preferentially inhabit the {\La} surface rather than the interface~\cite{Zhang-2010}.  Ideally, future theoretical work should examine the evolution of the electronic structure in the ``tail'' as a function of O$^{2-}$ vacancy density and location.  It also remains to be determined whether the absence of superconductivity from the ``tail'' region is merely due to a sub-critical carrier density, or if the $d_{xz,yz}$ orbital character also plays some role.  

At the maximum $n_{2D}$ which we are able to simulate (Fig.~4c), only the top TiO$_2$ monolayer at the interface still has a $d_{xy}$ character, with $d_{xz,yz}$ states dominating below.  We illustrate the effects of a back-gate electric field in Fig.~4d: as $V_g$ increases, the carrier density in the superconducting channel rises and a shift to the overdoped side of the superconducting dome occurs (as seen in Fig.~2c).  In parallel, electrons in the ``tail'' decompress away from the interface due to band-bending from the electric field, migrating hundreds of nanometres into the bulk.  This migration creates the 3D FS responsible for the SdH oscillations which we observe with $H/\!/[001]$, whose frequency scales with the total carrier density.  Between the interface and the bulk, the carrier density of the $d_{xz,yz}$-dominated region remains roughly constant: electrons which it ``loses'' to deeper-lying bulk states are replaced by electrons from the interface.  The presence of a large carrier population below the interface results in a screening of the electric field, thus explaining the relatively small increase of $d$ to 19~nm at $V_g=350$~V compared to $d>40$~nm reported at much smaller back-gate fields in the literature~\cite{Shalom-2010-2}.  Finally, Figs.~4e,f display exaggerated sketches illustrating the evolution of the FS as we move deeper into the {\Sr}, from $d_{xz,yz}$ domination (Fig.~4e) to a gradual recovery of $d_{xy,xz,yz}$ degeneracy (Fig.~4f) over a lengthscale $\geq$~120~nm.  While the microscopic origins of this long-distance evolution are still unclear, our work shows that functional oxide devices can reliably hope to profit from a renormalised electronic structure tens of nanometres away from a symmetry-breaking interface.  

\section*{Methods}
\footnotesize{\textit{
Two series of {\La}/{\Sr} heterostructures, ``A'' and ``B'', were grown using a standard pulsed laser deposition system manufactured by Twente Solid State Technology B.V., equipped with a reflection high-energy electron diffraction (RHEED) facility.  We use 0.5mm thick commercial 5$\times$5~mm {\Sr} (001) ``STEP'' substrates from Shinkosha: these are HF-treated for TiO$_2$ termination and cleaned by the manufacturer, then vacuum-packed for shipping.  We do not perform any additional surface cleaning or annealing prior to deposition: the substrates are loaded directly into our PLD chamber, which is subsequently evacuated to base pressure ($\lesssim$~10$^{-8}$~mbar) prior to backfilling with 10$^{-3}$~mbar O$_2$.  The substrate is then heated to growth temperature (800$^\circ$C).  Series A and B both feature 10 unit cells of {\La}, deposited using a total incident laser energy of 9~mJ focussed onto a 6~mm$^2$ rectangular spot.  The O$_2$ pressure and substrate temperature were maintained at 10$^{-3}$~mbar and 800$^\circ$C respectively for both sample series throughout the deposition process.  Subsequently, A-type samples underwent an annealing stage: after cooling to 500$^\circ$C at 10$^{-3}$~mbar, the O$_2$ pressure was increased to 0.1~bar.  The temperature was held at 500$^\circ$C for 30 minutes before natural cooling to 20$^\circ$C in 0.1~bar O$_2$.  In contrast, B-type samples were cooled naturally to 20$^\circ$C in 10$^{-3}$~mbar O$_2$.}} 
 
\footnotesize{\textit{
To fabricate Hall bars on these {\La}/{\Sr} films, we first defined contact pad areas using photolithography with AZ5214 photoresist.  2~nm Ti followed by 8~nm Au were evaporated directly onto the {\La} surface; the remaining photoresist was removed by soaking in acetone for 30 minutes, then rinsed in IPA.  Sample B also had an Au-Ti back gate deposited across the entire base of the {\Sr} substrate prior to fabrication.  The Hall bars were defined using a similar photolithography process and the Hall bar mesas etched using a dry Ar ion technique (at a slow rate of 1~\AA~s$^{-1}$ to avoid any substrate heating).  The Hall bar width was 80~$\mu$m and the voltage contact separation 660~$\mu$m.  Multiple Hall bars were fabricated on each 5$\times$5~mm substrate: tests show that the Hall bars are electrically isolated from each other (thus ruling out any parasitic conduction from the {\Sr} surface) and display similar transport properties (indicating that our heterostructures are homogeneous).  Prior to measurement, the Hall bars were mounted in thermally-conductive chip-carriers, with electrical contacts made using 10~$\mu$m Au wires ball-bonded to the Au-Ti contact pads.}}  

\footnotesize{\textit{
Transport data were acquired in a cryogen-free dilution refrigerator, using an AC technique with two digital lock-in amplifiers and a current source outputting 500~nA at 19~Hz.  This value was chosen to maximise the signal-to-noise ratio whilst minimising sample heating below 0.1~K.  Our noise threshold is approximately 1~nV.  The substrate capacitance was measured with femtoFarad sensitivity for gate voltages up to 500~V using a General Radio 1621 manual capacitance bridge.  All results presented in this work were qualitatively reproducible over a 6-month period comprising numerous cool-downs of both samples.  A total of 6 ``A-type'' and 4 ``B-type'' heterostructures were fabricated in our laboratory using identical ``recipes'' to those detailed above: all samples displayed similar behaviour to those discussed in the present work.}}

\section*{Acknowledgements}

The authors gratefully acknowledge discussions with H. Hilgenkamp, A. Fujimori and I. Martin.  This work was supported by the National Research Foundation, Singapore, through Grant NRF-CRP4-2008-04.  The research at the University of Nebraska-Lincoln (UNL) was supported by the National Science Foundation through the Materials Research Science and Engineering Center (Grant No. DMR-0820521) and the Designing Materials to Revolutionize and Engineer our Future (DMREF) Program (Grant No. DMR-1234096). Computations were performed at the UNL Holland Computing Center and the Center for Nanophase Materials Sciences, which is sponsored at Oak Ridge National Laboratory by the Scientific User Facilities Division, Office of Basic Energy Sciences, U.S. Department of Energy.

\section*{Author contributions}

A.P.P. and C.P. conceived the project.  A.D. and T.W. grew the heterostructures.  K.L., S.H. and C.B. fabricated and tested the Hall bars.  A.P.P. and A.P. set up and performed the experiments.  T.P. and E.T. contributed the band structure and Fermi surface calculations.  A.P.P. and C.P. wrote the paper.  C.P. supervised the entire study.  All authors discussed the results and manuscript.  
\\
\section*{Competing Interests}

The authors declare that they have no competing financial interests.


\begin{thebibliography}{10}

\bibitem{Ohtomo-2004}
Ohtomo, A. and Hwang, H.Y.
\newblock A high-mobility electron gas at the LaAlO$_3$/SrTiO$_3$
  heterointerface.
\newblock \emph{Nature} \textbf{427}, 423 (2004).

\bibitem{Brinkman-2007}
Brinkman, A. et~al.
\newblock Magnetic effects at the interface between non-magnetic oxides.
\newblock \emph{Nature Mater.} \textbf{6}, 493 (2007).

\bibitem{Reyren-2007}
Reyren, N. et~al.
\newblock Superconducting Interfaces Between Insulating Oxides.
\newblock \emph{Science} \textbf{317}, 1196 (2007).

\bibitem{Salluzzo-2009}
Salluzzo, M. et~al.
\newblock Orbital Reconstruction and the Two-Dimensional Electron Gas at the
  LaAlO$_3$/SrTiO$_3$ Interface.
\newblock \emph{Phys. Rev. Lett.} \textbf{102}, 166804 (2009).

\bibitem{Caviglia-2008}
Caviglia, A.D. et~al.
\newblock Electric field control of the LaAlO$_3$/SrTiO$_3$ interface ground
  state.
\newblock \emph{Nature} \textbf{456}, 624 (2008).

\bibitem{Bell-2009}
Bell, C. et~al.
\newblock Dominant Mobility Modulation by the Electric Field Effect at the
  LaAlO$_3$/SrTiO$_3$ Interface.
\newblock \emph{Phys. Rev. Lett.} \textbf{103}, 226802 (2009).

\bibitem{Caviglia-2010}
Caviglia, A.D. et~al.
\newblock Two-Dimensional Quantum Oscillations of the Conductance at
  LaAlO$_3$/SrTiO$_3$ Interfaces.
\newblock \emph{Phys. Rev. Lett.} \textbf{105}, 236802 (2010).

\bibitem{Shalom-2010}
Ben~Shalom, M., Ron, A., Palevski, A., and Dagan, Y.
\newblock Shubnikov-De Haas Oscillations in SrTiO$_3$/LaAlO$_3$ Interface.
\newblock \emph{Phys. Rev. Lett.} \textbf{105}, 206401 (2010).

\bibitem{Chambers-2010}
Chambers, S.A. et~al.
\newblock Instability, intermixing and electronic structure at the epitaxial
  LaAlO$_3$/SrTiO$_3$(001) heterojunction.
\newblock \emph{Surf. Sci. Rep.} \textbf{65}, 317 (2011).

\bibitem{Popovic-2008}
Popovi{\'c}, Z., Satpathy, S., and Martin, R.M.
\newblock Origin of the Two-Dimensional Electron Gas at the LaAlO$_3$ on
  SrTiO$_3$ interface.
\newblock \emph{Phys. Rev. Lett.} \textbf{101}, 256801 (2008).

\bibitem{Pentcheva-2010}
Pentcheva, R. et~al.
\newblock Parallel Electron-Hole Bilayer Conductivity from Electronic Interface
  Reconstruction.
\newblock \emph{Phys. Rev. Lett.} \textbf{104}, 166804 (2010).

\bibitem{Delugas-2011}
Delugas, P. et~al.
\newblock Spontaneous 2-Dimensional Carrier Confinement at the n-Type
  SrTiO$_3$/LaAlO$_3$ Interface.
\newblock \emph{Phys. Rev. Lett.} \textbf{106}, 166807 (2011).

\bibitem{Khalsa-2012}
Khalsa, G. and MacDonald, A.H.
\newblock Theory of the SrTiO$_3$ surface state two-dimensional electron gas.
\newblock \emph{Phys. Rev. B} \textbf{86}, 125121 (2012).

\bibitem{Santander-2011}
Santander-Syro, A.F. et~al.
\newblock Two-dimensional electron gas with universal subbands at the surface
  of SrTiO$_3¨$.
\newblock \emph{Nature} \textbf{469}, 189 (2011).

\bibitem{Marel-2011}
van~der Marel, D., van Mechelen, J.L.M., and Mazin, I.I.
\newblock Common Fermi-liquid origin of $T^2$ resistivity and superconductivity
  in $n$-type SrTiO$_3$.
\newblock \emph{Phys. Rev. B} \textbf{84}, 205111 (2011).

\bibitem{Herranz-2007}
Herranz, G. et~al.
\newblock High Mobility in LaAlO$_3$/SrTiO$_3$ Heterostructures: Origin,
  Dimensionality, and Perspectives.
\newblock \emph{Phys. Rev. Lett.} \textbf{98}, 216803 (2007).

\bibitem{Basletic-2008}
Basleti{\'c}, M. et~al.
\newblock Mapping the spatial distribution of charge carriers in LaAlO$_3$
  /SrTiO$_3$ heterostructures.
\newblock \emph{Nature Mater.} \textbf{7}, 621 (2008).

\bibitem{Nakagawa-2006}
Nakagawa, N., Hwang, H.Y., and Muller, D.
\newblock Why some interfaces cannot be sharp.
\newblock \emph{Nature Mater.} \textbf{5}, 204 (2006).

\bibitem{Pavlenko-2012-2}
Pavlenko, N., Kopp, T., Tsymbal, E.Y., Mannhart, J., and Sawatzky, G.A.
\newblock Oxygen vacancies at titanate interfaces: Two-dimensional magnetism
  and orbital reconstruction.
\newblock \emph{Phys. Rev. B} \textbf{86}, 064431 (2012).

\bibitem{Petrovic-2013}
Petrovi{\'c}, A.P. et~al.
\newblock The Vortex Signature of Discrete Ferromagnetic Dipoles at the
  LaAlO$_3$/SrTiO$_3$ Interface.
\newblock \emph{arXiv:1311.2323}  (2013).

\bibitem{Son-2010}
Son, J. et~al.
\newblock Epitaxial SrTiO$_3$ films with electron mobilities exceeding
  30000~cm$^2$V$^{-1}$s$^{-1}$.
\newblock \emph{Nature Mater.} \textbf{9}, 482 (2010).

\bibitem{Joshua-2011}
Joshua, A., Pecker, S., Ruhman, J., Altman, E., and Ilani, S.
\newblock A Universal Critical Density Underlying the Physics of Electrons at
  the LaAlO$_3$/SrTiO$_3$ Interface.
\newblock \emph{Nat. Commun.} \textbf{3}, 1129 (2012).

\bibitem{Koonce-1967}
Koonce, C.S., Cohen, M.L., Schooley, J.F., Hosler, W.R., and Pfeiffer, E.R.
\newblock Superconducting Transition Temperatures of Semiconducting SrTiO$_3$.
\newblock \emph{Phys. Rev.} \textbf{163}, 380 (1967).

\bibitem{Lin-2013}
Lin, X., Zhu, Z., Fauqu{\'e}, B., and Behnia, K.
\newblock Fermi Surface of the Most Dilute Superconductor.
\newblock \emph{Phys. Rev. X} \textbf{3}, 021002 (2013).

\bibitem{Gregory-1979}
Gregory, B., Arthur, J., and Seidel, G.
\newblock Measurements of the Fermi surface of SrTiO$_3$: Nb.
\newblock \emph{Phys. Rev. B} \textbf{19}, 1039 (1979).

\bibitem{Kim-2011}
Kim, M. et~al.
\newblock Fermi Surface and Superconductivity in Low-Density High-Mobility
  $\delta$-Doped SrTiO$_3$.
\newblock \emph{Phys. Rev. Lett.} \textbf{107}, 106801 (2011).

\bibitem{Mattheiss-1972}
Mattheiss, L.F.
\newblock Effect of the 110K Phase Transition on the SrTiO$_3$ Conduction
  Bands.
\newblock \emph{Phys. Rev. B} \textbf{6}, 4740 (1972).

\bibitem{Bark-2011}
Bark, C.W. et~al.
\newblock Tailoring a two-dimensional electron gas at the LaAlO$_3$/SrTiO$_3$
  (001) interface by epitaxial strain.
\newblock \emph{Proc. Natl. Acad. Sci. U.S.A.} \textbf{108}, 4720 (2011).

\bibitem{Honig-2013}
Honig, M. et~al.
\newblock Local electrostatic imaging of striped domain order in
  LaAlO$_3$/SrTiO$_3$.
\newblock \emph{Nature Mater.} \textbf{12}, 1112 (2013).

\bibitem{Allen-2013}
Allen, S.J. et~al.
\newblock Conduction-band edge and Shubnikov-de Haas effect in
  low-electron-density SrTiO$_3$.
\newblock \emph{Phys. Rev. B} \textbf{88}, 045114 (2013).

\bibitem{Pentcheva-2006}
Pentcheva, R. and Pickett, W.E.
\newblock Charge localization or itineracy at LaAlO$_3$/SrTiO$_3$ interfaces:
  Hole polarons, oxygen vacancies, and mobile electrons.
\newblock \emph{Phys. Rev. B} \textbf{74}, 035112 (2006).

\bibitem{Huijben-2013}
Huijben, M. et~al.
\newblock Defect Engineering in Oxide Heterostructures by Enhanced Oxygen
  Surface Exchange.
\newblock \emph{Adv. Funct. Mater.} \textbf{23}, 5240 (2013).

\bibitem{Zhang-2010}
Zhang, L. et~al.
\newblock Origin of insulating behavior of the $p$-type LaAlO$_3$/SrTiO$_3$
  interface: Polarization-induced asymmetric distribution of oxygen vacancies.
\newblock \emph{Phys. Rev. B} \textbf{82}, 125412 (2010).

\bibitem{Shalom-2010-2}
Ben~Shalom, M., Sachs, M., Rakhmilevitch, D., Palevski, A., and Dagan, Y.
\newblock Tuning Spin-Orbit Coupling and Superconductivity at the
  SrTiO$_3$/LaAlO$_3$ Interface: A Magnetotransport Study.
\newblock \emph{Phys. Rev. Lett.} \textbf{104}, 126802 (2010).

\end{thebibliography}
\end{document}